\documentclass[prc,superscriptaddress,preprint,showpacs,showkeys,aps]{revtex4}
\usepackage[T1]{fontenc}
\usepackage[latin1]{inputenc}
\usepackage{graphicx}
\usepackage{array}


\begin{document}

\def\deeps{\mbox{D($e,e^\prime p_s$)}}
\def\dee{\mbox{D($e,e^\prime$)}}
\def\f2neff{$F_{2n}^{\mbox{\tiny eff}}$}

%
%
%
%




\newcommand*{\ASU}{Arizona State University, Tempe, Arizona 85287-1504}
\affiliation{\ASU}
\newcommand*{\UCLA}{University of California at Los Angeles, Los Angeles, California  90095-1547}
\affiliation{\UCLA}
\newcommand*{\CMU}{Carnegie Mellon University, Pittsburgh, Pennsylvania 15213}
\affiliation{\CMU}
\newcommand*{\CUA}{Catholic University of America, Washington, D.C. 20064}
\affiliation{\CUA}
\newcommand*{\SACLAY}{CEA-Saclay, Service de Physique Nucl\'eaire, F91191 Gif-sur-Yvette,Cedex, France}
\affiliation{\SACLAY}
\newcommand*{\CNU}{Christopher Newport University, Newport News, Virginia 23606}
\affiliation{\CNU}
\newcommand*{\UCONN}{University of Connecticut, Storrs, Connecticut 06269}
\affiliation{\UCONN}
\newcommand*{\ECOSSEE}{Edinburgh University, Edinburgh EH9 3JZ, United Kingdom}
\affiliation{\ECOSSEE}
\newcommand*{\EMMY}{Emmy-Noether Foundation, Germany}
\affiliation{\EMMY}
\newcommand*{\FIU}{Florida International University, Miami, Florida 33199}
\affiliation{\FIU}
\newcommand*{\FSU}{Florida State University, Tallahassee, Florida 32306}
\affiliation{\FSU}
\newcommand*{\GWU}{The George Washington University, Washington, DC 20052}
\affiliation{\GWU}
\newcommand*{\ECOSSEG}{University of Glasgow, Glasgow G12 8QQ, United Kingdom}
\affiliation{\ECOSSEG}
\newcommand*{\ISU}{Idaho State University, Pocatello, Idaho 83209}
\affiliation{\ISU}
\newcommand*{\INFNFR}{INFN, Laboratori Nazionali di Frascati, Frascati, Italy}
\affiliation{\INFNFR}
\newcommand*{\INFNGE}{INFN, Sezione di Genova, 16146 Genova, Italy}
\affiliation{\INFNGE}
\newcommand*{\ORSAY}{Institut de Physique Nucleaire ORSAY, Orsay, France}
\affiliation{\ORSAY}
\newcommand*{\BONN}{Institute f\"{u}r Strahlen und Kernphysik, Universit\"{a}t Bonn, Germany}
\affiliation{\BONN}
\newcommand*{\ITEP}{Institute of Theoretical and Experimental Physics, Moscow, 117259, Russia}
\affiliation{\ITEP}
\newcommand*{\JMU}{James Madison University, Harrisonburg, Virginia 22807}
\affiliation{\JMU}
\newcommand*{\KYUNGPOOK}{Kyungpook National University, Daegu 702-701, South Korea}
\affiliation{\KYUNGPOOK}
\newcommand*{\MIT}{Massachusetts Institute of Technology, Cambridge, Massachusetts  02139-4307}
\affiliation{\MIT}
\newcommand*{\UMASS}{University of Massachusetts, Amherst, Massachusetts  01003}
\affiliation{\UMASS}
\newcommand*{\MOSCOW}{Moscow State University, General Nuclear Physics Institute, 119899 Moscow, Russia}
\affiliation{\MOSCOW}
\newcommand*{\UNH}{University of New Hampshire, Durham, New Hampshire 03824-3568}
\affiliation{\UNH}
\newcommand*{\NSU}{Norfolk State University, Norfolk, Virginia 23504}
\affiliation{\NSU}
\newcommand*{\OHIOU}{Ohio University, Athens, Ohio  45701}
\affiliation{\OHIOU}
\newcommand*{\ODU}{Old Dominion University, Norfolk, Virginia 23529}
\affiliation{\ODU}
\newcommand*{\PITT}{University of Pittsburgh, Pittsburgh, Pennsylvania 15260}
\affiliation{\PITT}
\newcommand*{\RPI}{Rensselaer Polytechnic Institute, Troy, New York 12180-3590}
\affiliation{\RPI}
\newcommand*{\RICE}{Rice University, Houston, Texas 77005-1892}
\affiliation{\RICE}
\newcommand*{\Turkey}{Sakarya University, Sakarya, Turkey}
\affiliation{\Turkey}
\newcommand*{\URICH}{University of Richmond, Richmond, Virginia 23173}
\affiliation{\URICH}
\newcommand*{\SCAROLINA}{University of South Carolina, Columbia, South Carolina 29208}
\affiliation{\SCAROLINA}
\newcommand*{\JLAB}{Thomas Jefferson National Accelerator Facility, Newport News, Virginia 23606}
\affiliation{\JLAB}
\newcommand*{\UNIONC}{Union College, Schenectady, NY 12308}
\affiliation{\UNIONC}
\newcommand*{\NONE}{unknown}
\newcommand*{\VT}{Virginia Polytechnic Institute and State University, Blacksburg, Virginia   24061-0435}
\affiliation{\VT}
\newcommand*{\VIRGINIA}{University of Virginia, Charlottesville, Virginia 22901}
\affiliation{\VIRGINIA}
\newcommand*{\WM}{College of William and Mary, Williamsburg, Virginia 23187-8795}
\affiliation{\WM}
\newcommand*{\YEREVAN}{Yerevan Physics Institute, 375036 Yerevan, Armenia}
\affiliation{\YEREVAN}
\newcommand*{\NOWOHIOU}{Ohio University, Athens, Ohio  45701}
\newcommand*{\NOWUNH}{University of New Hampshire, Durham, New Hampshire 03824-3568}
\newcommand*{\NOWUMASS}{University of Massachusetts, Amherst, Massachusetts  01003}
\newcommand*{\NOWMIT}{Massachusetts Institute of Technology, Cambridge, Massachusetts  02139-4307}
\newcommand*{\NOWURICH}{University of Richmond, Richmond, Virginia 23173}
\newcommand*{\NOWODU}{Old Dominion University, Norfolk, Virginia 23529}
\newcommand*{\NOWGEISSEN}{Physikalisches Institut der Universitaet Giessen, 35392 Giessen, Germany}
\newcommand*{\NOWNONE}{unknown, }
\newcommand*{\NOWLANL }{ Los Alamos National Laboratory, Los Alamos, New Mexico 87545}


\author{A.V.~Klimenko}
     \email{klimenko@lanl.gov}
     \altaffiliation[Current address:]{\NOWLANL}
     \affiliation{\ODU}
\author{S.E.~Kuhn}
     \email{skuhn@odu.edu}
     \affiliation{\ODU}
\author{C.~Butuceanu}
      \affiliation{\WM}
\author{K.S.~Egiyan}
      \affiliation{\YEREVAN} 
\author{K.A.~Griffioen}
     \affiliation{\WM}
\author {G.~Adams} 
\affiliation{\RPI}
\author {P.~Ambrozewicz} 
\affiliation{\FIU}
\author {M.~Anghinolfi} 
\affiliation{\INFNGE}
\author {G.~Asryan} 
\affiliation{\YEREVAN}
\author {H.~Avakian} 
\affiliation{\JLAB}
% ...
 
\collaboration{The CLAS Collaboration}
     \noaffiliation

\date{\today}

\title{Electron Scattering From High-Momentum Neutrons In Deuterium}

\begin{abstract}
We report results from an experiment measuring the semi-inclusive reaction \deeps\   where the 
proton $p_s$ is moving at a large angle relative to the momentum transfer. If we assume that the 
proton was a spectator to the reaction taking place on the neutron in deuterium, the initial state of 
that neutron can be inferred. This method, known as spectator tagging, can be used to study electron 
scattering from high-momentum (off-shell) neutrons in deuterium. The data were taken with a 5.765 GeV 
electron beam on a deuterium target in Jefferson Laboratory's Hall B, using the CLAS detector. 
A reduced cross section was extracted for different values of final-state missing mass $W^{*}$, 
backward proton momentum $\vec{p}_{s}$ and momentum transfer $Q^{2}$. The data are 
compared to a simple PWIA spectator model.  A strong enhancement in the data observed at 
transverse kinematics is not reproduced by the PWIA model. This enhancement can likely
be associated with the contribution of final state interactions (FSI) that were not incorporated into the 
model. A ``bound neutron structure function'' \f2neff  was extracted   as a function of $W^{*}$ and the 
scaling variable $x^{*}$ at extreme backward kinematics, where effects of FSI appear to be smaller. 
For $p_{s}>400\,\,\mathrm{MeV/c}$, where the neutron is far off-shell, the model overestimates the 
value of \f2neff in the region of $x^{*}$ between 0.25 and 0.6. A modification of the bound neutron 
structure function is one of possible effects that can cause the observed deviation.
\end{abstract}
\maketitle

\section{Introduction}

Decades before the 
nucleon 
substructure
was discovered, numerous models were developed that successfully describe
most nuclear phenomena only in terms of nucleons, their excited states
and strong force mediators - mesons. Nucleons and mesons are often
called the {}``conventional'' degrees of freedom of nuclear physics.
The fundamental theory of strong interactions, quantum chromodynamics (QCD),
describes physical processes in terms of quarks and gluons. QCD is
very successful in describing the interaction of quarks at short distances,
where perturbative methods, similar to those of quantum electrodynamics
(QED) in atomic physics, are applicable. However, the same perturbative
methods cannot be applied to solve QCD at the length scales of a nucleus.
The present difficulty to make rigorous predictions based on QCD at
low momenta (corresponding to large distance scales)
leaves us no choice but to continue to employ nuclear theories based
on {}``effective'' degrees of freedom - nucleons and mesons. In
an attempt to resolve this discontinuity of theories, the focus of
modern nuclear physics has turned to the intermediate region where QCD is
not yet solvable, but the quark-gluon substructure of the nucleons
must be taken into account in the
nuclear models. 

One example of the interface between a hadronic and a quark-based
description is the (possible) modification of the (quark--) structure
of a nucleon that is part of a tightly bound pair. Due to the
Heisenberg uncertainty principle, large momenta of the nucleons inside
the nucleus can be associated with small internucleon spatial separations. 
The kinematical conditions are particularly clean in the case of the deuteron,
where the relative motion of the two nucleons is completely described
by the wave function in momentum space, $\psi(p)$. In all models
of the deuterium nucleus, the nucleons have mostly low momenta and
therefore are relatively far apart. However, even in the wave functions
obtained from non-relativistic models of the nucleon-nucleon potential,
there is a probability for the nucleons to have momenta high enough
so that the proton and neutron can come very close together or even overlap.
In such high density configurations the quark distribution within
a nucleon can become modified either through off-shell effects \cite{Melnitchouk:1996vp}
or through direct modification of 
the shape and size of the nucleon \cite{Frankfurt:1981mk,Close:1984zn}.
It is also possible that under these conditions the nucleons start
to exchange quarks with each other or even merge into a single {}``six-quark
bag'' \cite{Carlson:1994ga,Carlson:1999uk}. The quark-gluon degrees
of freedom thus might play a direct role in modifying nucleon
structure in high-density nuclear configurations. The analysis presented
here is aimed at advancing the understanding of high density, high
momentum nuclear matter.

To study these high density configurations, we have used electron
scattering from a high-momentum nucleon within a nucleus. In the case
of a deuteron target this can be easily verified by taking advantage
of the inherently simple structure of the two-nucleon system. If all
the momentum and energy is transferred to the neutron, the proton is
a spectator to the reaction and recoils with its initial momentum.
Assuming that the detected proton was indeed a spectator to the reaction,
the initial momentum of the struck neutron can be obtained using momentum
conservation. Thus the neutron is {}``tagged'' by the backward going
spectator proton
(for a extensive discussion of the spectator picture
see, e.g., the papers by Simula~\cite{Simula:1996xk} and 
Meltnitchouk {\it et al.}~\cite{Melnitchouk:1996vp}).
Measurement of a high-momentum
proton emitted backwards relative to the momentum transfer direction
 allows us to infer that the electron interacted with a high-momentum
neutron in deuterium. 

\section{Theoretical Models}

\subsection{Nucleons in the Nuclear Medium}

\label{sec:Nucleons-In-a}

Energy conservation applied to the deuterium nucleus requires that
the total energy of the proton and neutron bound within a deuteron
equals the mass of the deuterium nucleus:

\begin{equation}
E_{p}+E_{n}=M_{d} .
\label{eq:mdeq}\end{equation}
At the same time, the mass of the deuteron is less than the mass of
a free proton plus the mass of a free neutron, $M_{d}=M_{p}+M_{n}-2.2246\,\,\,\mathrm{MeV}$.
Therefore, both the bound neutron and proton can not be on the mass shell
at the same time. In the {}``instant form'' dynamics, one of the
nucleons is assumed to be on-shell, while the other one is off-shell
and its off-shell energy is $E_{n}^{*}=M_{d}-\sqrt{M_{p}^{2}+p_{s}^{2}}$. 

The final state motion of the on-shell ({}``spectator'') nucleon
can be described by its momentum $\vec{p}_{s}$ or the light cone
fraction $\alpha_{s}$:
\begin{equation}
\alpha_{s}=\frac{E_{s}-p_{s_{||}}}{M} ,
\label{eq:alphas}\end{equation}
where $p_{s}^{\mu}=(E_{s},\vec{p}_{T}, p_{s_{||}})$ is the spectator
proton momentum 4-vector. The component $p_{s_{||}}$ of the proton
momentum is in the direction of the momentum transfer $\hat{q}$, and
$\vec{p}_{T}$ is transverse to $\hat{q}$.

Using a non-relativistic wave function $\psi_{NR}(p_{s})$,
the ``target density'' of neutrons which are correlated with
 spectator protons of
momentum $\vec{p}_{s}$ can be expressed as:
\begin{equation}
P(\vec{p}_{s})=J\cdot|\psi_{NR}(p_{s})|^{2} ,
\label{eq:pps}\end{equation}
\noindent where $J=1+\frac{p_{s_{||}}}{E_{n}^{*}}=\frac{(2-\alpha_{s})M_{D}}{2(M_{D}-E_{s})}$
is a flux factor that accounts for the motion of the struck nucleon.

The probability $P(\vec{p}_{s})$ is related to the spectral function:
\begin{equation}
S(\alpha_{s},p_{T})\frac{d\alpha_{s}}{\alpha_{s}}d^{2}p_{T}=P(\vec{p}_{s})d^{3}p_{s} ,
\label{eq:salphas}\end{equation}
\noindent which yields $S=E_{s}\cdot P(\vec{p}_{s})$.

In the light-cone dynamics framework, a non-relativistic deuterium
wave function can be rescaled to account for relativistic effects
at high momenta \cite{Frankfurt:1981mk}:
\begin{equation}
S^{LC}(\alpha_{s},p_{T})\frac{d\alpha_{s}}{\alpha_{s}}d^{2}p_{T}=|\psi_{NR}(|\vec{k}|^{2})|^{2}d^{3}k
\label{eq:sapt}\end{equation}
\begin{equation}
\alpha_{s}=1-\frac{k_{||}}{\sqrt{M^{2}+\vec{k}^{2}}}\end{equation}
\begin{equation}
\begin{array}{cc}
\vec{p}_{T}=\vec{k}_{T} & \,\,\,\,\,\,\,\,\,\,\,\,\,\,\, |\vec{k}| 
=\sqrt{\frac{M^{2}+ p_{T}^{2}}{\alpha_{s}(2-\alpha_{s})}-M^{2}} ,
\end{array}\label{eq:pp}\end{equation}
\noindent where $\alpha_{s}$ is the light-cone fraction of the nucleus
carried by the spectator nucleon and $k^{\mu}=(k_{0},\vec{k}_{T},k_{||})$
is its internal momentum, with $k_{0}=\sqrt{M^{2}+\vec{k}^{2}}$.
The relativistic effect, in this picture, manifests itself in that
the measured momentum of the nucleon $p_{s_{||}}$ is rescaled in the
lab frame from the internal momentum $k_{||}$. The resulting deuterium
momentum distribution is given by the spectral function:
\begin{equation}
S^{LC}(\alpha_{s},p_{T})=\frac{\sqrt{M^{2}+\vec{k}^{2}}}{2-\alpha_{s}}|\psi_{NR}(|\vec{k}|)|^{2} .
\label{eq:SpFun}\end{equation}
The spectral function is normalized to satisfy the relation:
\begin{equation}
\int\int\int S^{LC}(\alpha_{s},p_{T})\frac{d\alpha_{s}}{\alpha_{s}}d^{2}p_{T}=1 .
\label{eq:ints}\end{equation}
 In the PWIA spectator approximation, the recoiling proton is on-shell
at the moment of interaction and receives no energy or momentum transfer,
so that its initial and final momenta in the lab are the same.
The differential cross-section on a moving nucleon (with kinematics defined by the spectator variables
$\alpha_{s},p_{T}$) can then be calculated 
as:
\begin{equation}
\begin{array}{cc}
\frac{d\sigma}{dx^{*}dQ^{2}}=\frac{4\pi\alpha_{EM}^{2}}{x^{*}Q^{4}}\left[\frac{y^{*2}}{2(1+R)}+(1-y^{*})+\frac{M^{*2}x^{*2}y^{*2}}{Q^{2}}\frac{1-R}{1+R}\right]\\
\times F_{2}(x^{*},\alpha_{s},p_{T},Q^{2})\cdot S(\alpha_{s},p_{T})\frac{d\alpha_{s}}{\alpha_{s}}d^{2}p_{T}\end{array} ,
\label{eq:ccoffshell}\end{equation}
where $S(\alpha_{s},p_{T})\frac{d\alpha_{s}}{\alpha_{s}}d^{2}p_{T}$ is the probability to find
a spectator with the given kinematics.
In this expression, $F_{2}(x^{*},\alpha_{s},p_{T},Q^{2})$ is the ``off--shell'' structure function
of the struck neutron 
and $R=\frac{\sigma_{L}}{\sigma_{T}}$
is the ratio between the longitudinal and transverse cross sections.
The asterisk is used for variables that have been
defined in a manifestly covariant way.
For instance, the Bjorken scaling variable $x=\frac{Q^{2}}{2M\nu}$
and the variable $y=\frac{\nu}{E}$ that are valid for the scattering
from a free nucleon at rest are replaced with their counterparts
for the scattering on a moving neutron inside the deuteron:
\begin{equation}
x^{*}=\frac{Q^{2}}{2p_{N}^{\mu}q^{\mu}}\approx\frac{Q^{2}}{2M\nu(2-\alpha_{s})}=\frac{x}{2-\alpha_{s}}
\label{eq:x_star}\end{equation}
\[
y^{*}=\frac{p_{N}^{\mu}q_{\mu}}{p_{N}^{\mu}k_{\mu}}\approx y , \]
where $q^{\mu}=(\nu,\vec{q})$ is the momentum transfer 4-vector,
$k^{\mu}=(E,0,0,E)$ is the momentum 4-vector of the incident electron,
$p_{N}^{\mu}=(M_{d}-E_{s},-\vec{p}_{s})$ is the momentum 4-vector
of the off-shell neutron and $M_d$ is the mass of the deuterium nucleus.
In this approximation the struck nucleon is assumed to be on the energy
shell, but off its mass shell. The mass of the free nucleon $M$ is
therefore replaced with the off-shell mass of the bound nucleon:
\begin{equation}
M^{*2}=(M_{d}-E_{s})^{2}-\vec{p}_{s}^{\,2} .
\label{eq:m_star}\end{equation}
The invariant mass of the final hadronic state in \deeps $X$
scattering can be expressed as:
\begin{equation}
\begin{array}{cc}
W^{*2}=(p_{n}^{\mu}+q^{\mu})^{2}=M^{*2}-Q^{2}+2(M_D -E_{s})\nu+2p_{s_{||}}|\vec{q}|\\
=M^{^{*}2}-Q^{2}+2M\nu\left(2-\frac{E_{s}-p_{s_{||}}(|\vec{q}|/\nu)}{M}\right)\end{array} ,
\label{eq:wstar}\end{equation}
where 
it was assumed that $M_{d}\approx2M$. In the (Bjorken) limit of $|\vec{q}|/\nu\rightarrow1$
the fraction in the brackets of the last term in equation (\ref{eq:wstar})
takes the familiar form of the light-cone fraction of the nucleus
carried by the spectator proton $\alpha_{s}=\frac{E_{s}-p_{s_{||}}}{M}$,
yielding:
\begin{equation}
W^{*2}\approx M^{*2}-Q^{2}+2M\nu\left(2-\alpha_{s}\right) .
\label{eq:star}\end{equation}

If one assumes that $F_{2}$ is equal to its on-shell form,
$F_{2}(x^{*},\alpha_{s},p_{T},Q^{2}) = F_2^{free}(x^{*},Q^{2})$, and integrates
over the spectator kinematics, one obtains the usual convolution
result for the inclusive nuclear structure function $F_{2A}$. In
this picture the nucleus is built 
from free nucleons, i.e.
the struck nucleon has the same quark distribution
as a free nucleon. Any observed modification of the cross section from
that of a collection of free nucleons is just due to the kinematic rescaling (Eqs.~\ref{eq:x_star})
because of the motion of the nucleons inside the nucleus.
However, the difference in the $x$ dependence of the inclusive deep
inelastic cross section for free and bound nucleons observed by the
European Muon Collaboration (known as the EMC-effect \cite{Aubert:1983xm}),
cannot be interpreted solely in terms of such a kinematic shift. 
A large number of models have been proposed to explain
the EMC-effect. A good review of this subject is
 given by Sargsian {\em et al.} in Ref. \cite{Sargsian:2002wc}. 

The most conservative approach assumes that any modification of the bound nucleon structure
function is solely due to the fact that the struck nucleon is off its mass shell ($E < M$); for
example see Ref.~\cite{Melnitchouk:1996vp}. Other models invoke a change of the nucleon size
and therefore a rescaling of the structure function with momentum transfer $Q^2$, as in
Ref.~\cite{Close:1984zn}. 
Frankfurt and 
Strikman~\cite{Frankfurt:1981mk} link the 
modifications to the structure function with a suppression of
small (point-like) valence configurations of a strongly bound nucleon.
The most unconventional attempt to explain 
the EMC-effect is that of Carlson and Lassila \cite{Carlson:1994ga,Carlson:1999uk}
where nucleons inside of a nucleus in its high-density configuration
are thought to merge and form multiquark states. For the case of deuterium,
as much as 5\% of the wave function would be in
a 6-quark state in this model. 
The cross section for backward proton production is then expressed as
a convolution of the distribution function for the valence quarks in
a 6-quark cluster $V_{i}^{(6)}$ and the fragmentation function for
the 5-quark residuum into a backward proton, $D_{p/5q}(z)\propto(1-z)^{3}$, with $z=\alpha/(2-x)$.

Although all of these models can describe at least some aspects of the EMC--effect, they predict considerably
different changes of the internal structure of deeply bound nucleons. These changes are masked in
inclusive measurements, where one averages over all bound nucleons, most of which are below the
Fermi surface. By selecting tightly bound nucleon pairs (with a fast backward going spectator as ``tag''), 
our experiment can study these possible modifications more directly.



\subsection{Final State Interactions}

\label{sec:Final-State-Interactions}

The PWIA picture described above has to be modified to include the
effect of final state interactions (FSI) and two-body currents (meson exchange
currents). According to existing models (see below), there are kinematic
regions where FSI are thought to be small, and other regions where
FSI are enhanced. Reliable models of FSI exist for nucleon-nucleon
rescattering~\cite{Frankfurt:1996xx}. 
In the resonant and deep inelastic region, the estimation
of FSI is a lot more challenging. FSI can be modeled by replacing
the spectral function in Eq.~\ref{eq:ccoffshell} with
a distorted one: $S^{FSI}(\alpha_{s,}\vec{p}_{T})$ . 

Melnitchouk, Sargsian and Strikman \cite{Melnitchouk:1996vp} use
the $eD\rightarrow e\, p\, n$ reaction as a first estimate
of FSI in electron scattering from the deuteron. This calculation
shows that for $\alpha_{s}>2-x$ and $\vec{p}_{T}$ close to zero
FSI are small. In this model $S^{FSI}$ is evaluated using a distorted
wave impulse approximation (DWIA). According to this paper, FSI effects
should not strongly depend on $x$, thus the ratios of the cross section
for different ranges in $x$ should be a good tool to look for the EMC-effect
in the semi-inclusive $eD\rightarrow e\, p\, X$ process. In the limit
of large $x$, FSI become much more important for heavier nuclei,
where rescattering hadrons produced in the elementary 
deep inelastic scattering (DIS) off the
short-range correlation are dynamically enhanced. Therefore, deuterium
targets, in the authors' opinion, provide the best way of studying the origin of
the EMC effect. 

A more recent publication by Cioffi {\it et al.}~ \cite{CiofidegliAtti:2003pb}
discusses backward proton production and FSI associated with DIS by
evaluating $S^{FSI}$ within a hadronization framework. The reinteraction
of the backward-going spectator protons with the debris formed in
a hadronization process is modeled using an effective cross section:
\begin{equation}
\sigma^{\mbox{\small eff}}=\sigma^{NN}+\sigma^{\pi N}(n_{M}+n_{G}) ,
\label{eq:FSIcioffi}
\end{equation}
where $\sigma^{NN}$ and $\sigma^{\pi N}$ are the total 
nucleon-nucleon and meson-nucleon cross sections, respectively, and $n_{M}$
and $n_{G}$ are the effective numbers of created mesons and radiated
gluons. The cross section asymptotically tends to exhibit a simple
logarithmic behavior. The magnitude of the effective reinteraction
cross section differs significantly for different models, especially
at angles of proton emission $\theta\sim90^{o}$. This kinematic region
is proposed by the authors as the best place to test various models
of hadronization. In contrast with the calculation discussed in the
beginning of the section, the model of \cite{CiofidegliAtti:2003pb}
predicts significant FSI for proton momenta $|\vec{p}_{s}|>250\,\,\mathrm{MeV/c}$
even at extreme backward angles.

\section{Existing Data Overview}

\label{sec:Existing-Data-Overview}

Few data exist on the semi-inclusive scattering of a lepton from
deuterium with a recoiling nucleon in the backward direction with respect
to the momentum transfer. The data published so far were taken using
either neutrino or antineutrino beams and had very low statistics
that do not allow detailed investigation of the cross sections of
interest. These experiments (see Berge and Efremenko \cite{Berge:1978ie,Efremenko:1980sc})
focused on measuring the momentum, energy, and angular distributions
of protons in the backward hemisphere relative to the beam line. Despite
the low statistics, a notable difference in the distributions for
backward and forward protons was observed. The data were shown to
agree well with a pair-correlation model in which the detected backward
proton is assumed to be a spectator to the reaction.

The cross section ratio $\sigma^{Fe}/\sigma^{D}$ measured by the
European Muon Collaboration \cite{Aubert:1983xm} (where $\sigma^{Fe}$
and $\sigma^{D}$ are cross sections per nucleon for iron and deuterium
respectively) showed deviations from unity (now known as the EMC-effect)
that could not be explained only in terms of nucleon Fermi motion.
That was the first evidence that the nuclear medium influences DIS
processes. It provided an indication that nuclear matter is getting
modified as its density increases. The effect was later confirmed
by data from SLAC \cite{Bodek:1983qn,Arnold:1983mw} and CERN \cite{Bari:1985ga}.

An independent measurement of the modification of the quark structure
of nuclei was later done at Fermilab \cite{Alde:1990im} using continuum
dimuon production in high-energy hadron collisions, known as the Drell-Yan
process \cite{Drell:1970wh}. The measurement has shown no nuclear
dependence in the production of the dimuon pairs in the region $0.1<x<0.3$,
and therefore, no modification of the antiquark sea in this range.
A number of models developed to explain the EMC-effect in terms of
strong enhancement of the pion cloud were ruled out by this experiment.

A recent polarization transfer measurement by Dieterich and Strauch
\cite{Dieterich:2000mu,Dieterich:2001sj,Strauch:2002wu,Ransome:2002vx}
in the $^{4}\mathrm{He}(\vec{e},e'\vec{p})^{3}\mathrm{H}$ reaction
suggested medium modification of the electromagnetic form factors
of the nucleon. The observed 10\% deviation from unity could only
be explained by supplementing the conventional nuclear description
with effects due to medium modification of the nucleon as calculated
by the QMC model \cite{Blunden:1996kc,Lu:1997mu}.

A model in which the neutron and proton form a single 6-quark cluster
was recently tested \cite{Carlson:1999uk} against old backward proton
production data from neutrino scattering on deuterium collected at
Fermilab \cite{Kafka:1983}. These data had sufficient acceptance
for backward protons but were not previously analyzed for this signal.
The proton spectrum from neutrino and antineutrino scattering from
deuterium, taken at CERN \cite{Matsinos:1989fm}, was also discussed.
The authors compared the momentum distribution of backward protons
with the prediction of a 6-quark cluster model. Predictions of the
model were shown to be in good agreement with the data, however,
the statistics of the data were not sufficient to study the dependence
on any other kinematic variables.

In summary, existing data on inelastic scattering off nuclei average over at least
some of the relevant kinematic variables ($x$, $Q^2$, and the momentum of
the struck nucleon) and are often limited in statistics. Only a more detailed
analysis of the dependence of the cross section on these variables can yield
clear distinctions between different models and theoretical descriptions of nucleons
bound in nuclei. The experiment on the reaction \deeps \  described here is the
first to collect sufficient statistics for this purpose.

\section{Experimental Setup}

\begin{figure}
\includegraphics[%
  scale=0.6]{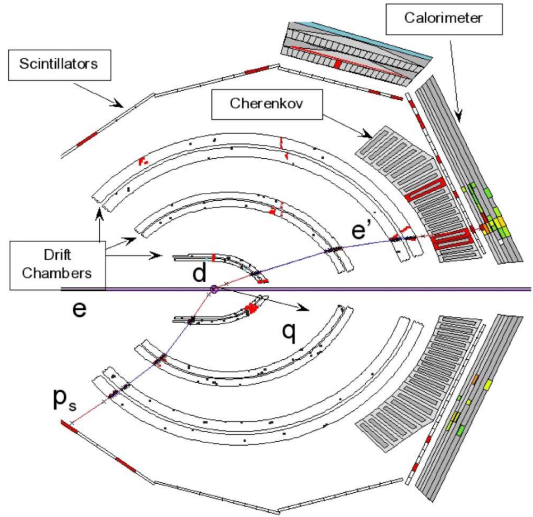}
\caption{\label{fig:CLAS} (Color online) CLAS event with forward electron detected in coincidence
with a backward proton.}
\end{figure}

The data were collected over a period of 46 calendar days in February
and March of 2002 at the Thomas Jefferson National Accelerator Facility
(TJNAF). We used a 5.75 GeV electron beam with an average
current of $6-9\,\,\mathrm{nA}$.
The experiment was staged in Hall B of the TJNAF, where the CEBAF
Large Acceptance Spectrometer (CLAS) is installed. Six superconducting
magnetic coils divide CLAS into six sectors symmetrically located
around the beamline. Each sector covers almost $60^{\circ}$ in azimuthal
angle and between $10^{\circ}$ and $140^{\circ}$ in polar angle, thus providing
almost $4\pi$ acceptance for charged particles. CLAS sectors are
equipped with identical sets of detector systems (Fig.~ \ref{fig:CLAS}):
1) three regions of drift chambers (DC) track charged particle's passage
though the region of magnetic field; 2) a layer of scintillating paddles
form the CLAS time-of-flight system (TOF); 3) the Cherenkov counters (CC)
are installed in the forward region ($10^{\circ}<\theta_{lab}<50^{\circ}$)
of the detector and efficiently discriminate electrons from pions
up to the particle momenta $p \approx2.7\,\,\mathrm{GeV/c}$; 4) several
layers of lead and scintillating paddles form the electromagnetic
calorimeter (EC) designed to separate electrons from minimum ionizing
particles. CLAS is described in detail in Ref. \cite{Mecking:2003zu}.

A conical cryogenic $5\,\,\mathrm{cm}$ target, installed in the center
of CLAS, was filled with liquid deuterium at a temperature of 22 K
and pressure of 1315 mbar with a density of $0.162\,\,\mathrm{g/cm^{3}}$.
The average beam current of 8 nA produced a luminosity
of $1.1\times10^{34}\,\,\mathrm{cm^{-2}\cdot s^{-1}}$. 

The CLAS trigger was formed by a coincidence between CC and EC. The
signal level for the trigger coincidence was set to be at least 1
photoelectron in CC and 0.5 GeV in EC. The level 2 trigger required
a DC track candidate in the sector of the calorimeter hit. With this
trigger configuration, the data rate was about $3\,\,\mathrm{kHz}$
and the dead time was usually less than 13\%.

\begin{figure}
\begin{flushleft}\includegraphics[%
  scale=0.24]{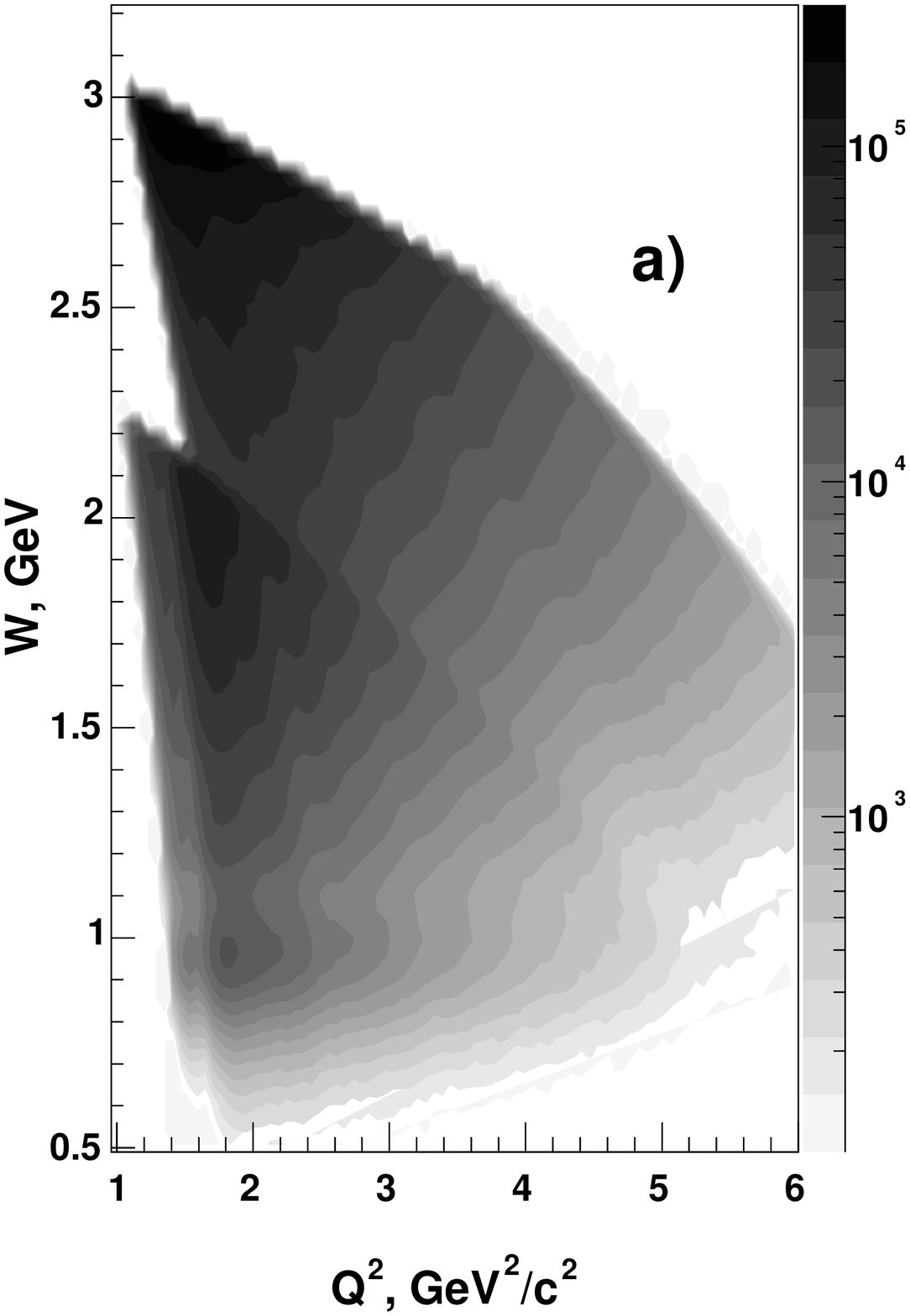}\includegraphics[%
  scale=0.24]{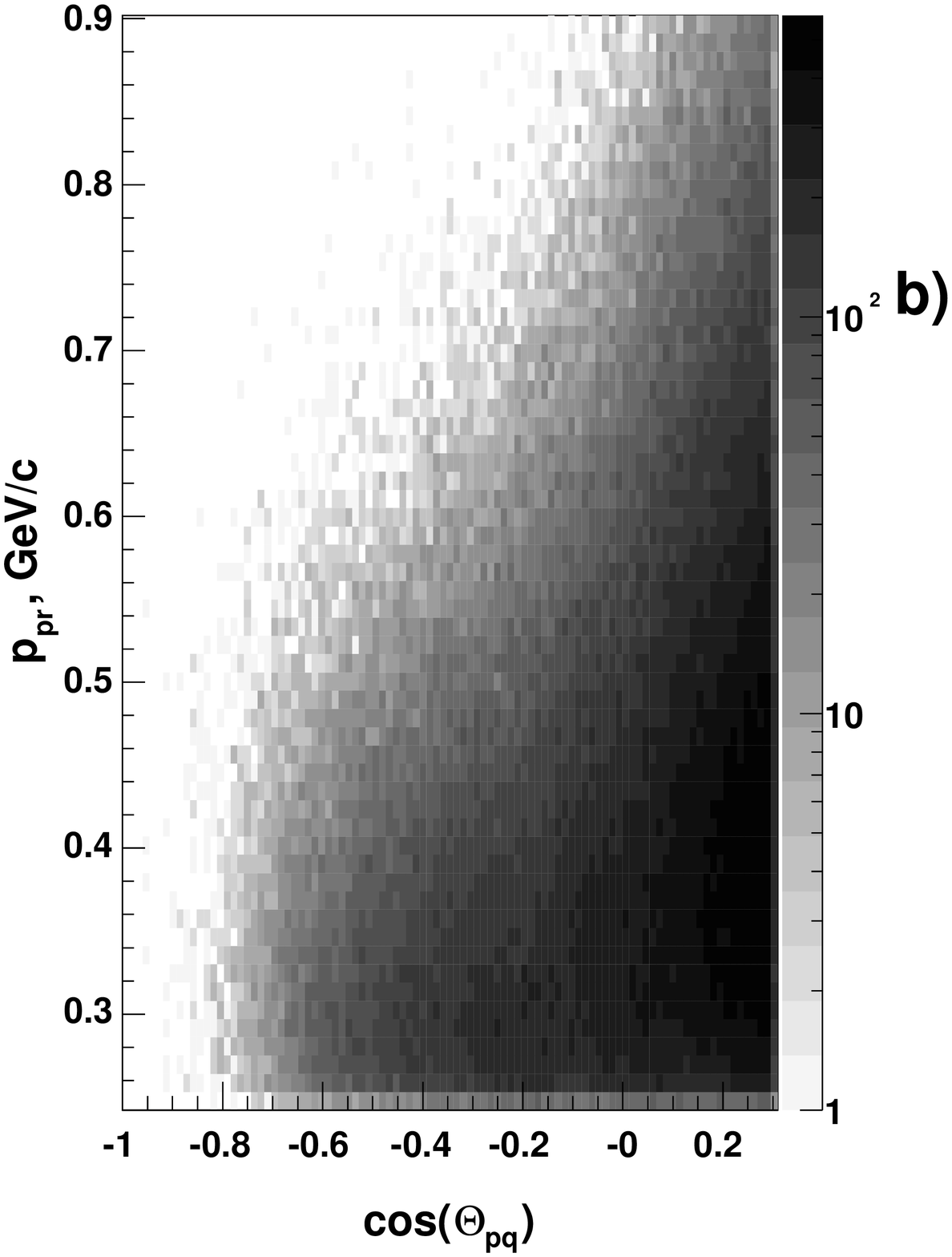}\end{flushleft}
\caption{\label{fig:kin_cov} Kinematic coverage 
for electrons ($W$ {\it vs.} $Q^2$)
 (a) and for recoiling protons 
(momentum $p_{pr}$ {\it vs.} polar angle $\theta_{pq}$) (b), within fiducial cuts.}
\end{figure}

Out of 4.5 billion events collected over the experimental run, only
350 thousand contain an electron in coincidence with a backward proton.
 The typical event of that type detected in
CLAS is shown in Fig. \ref{fig:CLAS}. The collected data sample has
wide coverage in kinematics of the electron and proton (Fig. \ref{fig:kin_cov}).
The momentum transfer $Q^{2}$ ranges between 1.2 and $5.5\,\,\mathrm{GeV^{2}/c^{2}}$,
while the invariant mass covers the quasi-elastic, resonant and deep
inelastic regions. Protons were detected at large angles relative
to the momentum transfer vector $\vec{q}$, 
 up to angles of $\theta_{pq}\approx145^{\circ}$ and with
momenta above $0.28$ GeV/c.

\section{Data Analysis}

In this section we discuss all the key analysis steps that led to
the extraction of the final results.

\subsection{Event Selection}

The focus of this
analysis is the $ed\rightarrow e'p_{s}X$ reaction, therefore events
containing coincidences between the scattered electron and recoiling proton
have to be selected first.

The scattered relativistic electron is expected to be the first particle
that arrives at the detectors after interacting with the target nucleus.
The particle was identified as an electron if it was the first in
the event and its charge was measured by the DC to be negative. Electron
identification (ID) cuts on the response of two of the remaining detector
systems, CC and EC, reduce the background of $\pi^{-}$ in the electron
spectrum. The CC are very efficient in pion rejection up to $P\approx2.7\,\,\mathrm{GeV/c}$,
where pions start to emit Cherenkov light. For lower momenta of the
particle $P<3.0$ GeV/c a software cut of 2.5 photoelectrons was required
to identify an electron. For the part of the data with particle momentum
$P>3.0$ GeV/c, a software cut of 1 photoelectron was used (and the fiducial
region increased - see below) to increase acceptance. The electron
produces an electromagnetic shower in the EC immediately after it
enters, while pions make mostly a minimum ionizing signal with a small
sampling fraction ($E/P$). The minimum ionizing particles can be
easily rejected by requiring that the visible energy deposited in the first
15 layers of the EC is 
$EC_{inner}>0.08\cdot P$ and the total visible energy in the EC is
$EC_{total}>0.22\cdot P$.

In order to reduce the systematic uncertainty in the quality of electron
identification, detector fiducial cuts are applied. The fiducial region
of CC is known to be within the limits of the EC fiducial region; therefore
only a CC cut needs to be applied. We defined the fiducial region
such that the CC was at least 90\% efficient.

In addition to the particle charge information, the DCs also measure
the length from the target to the TOF system
and the curvature of the track. From the curvature of the track
the particle momentum can be reconstructed. The proton is identified
using TOF time measurement ($t_{TOF}$) and DC momentum ($p_{DC}$)
and track length ($r$) information. Assuming a positively charged
particle is a proton, its velocity is given by
\begin{equation}
v_{DC}=\frac{p_{DC}}{\sqrt{p_{DC}^{2}+M_{p}^{2}}} ,
\label{eq:vdc}\end{equation}
where $M_{p}$ is proton mass. Then the time the proton travels from
the target to the TOF is $t_{DC}=r/v_{DC}$. The particle is identified
as a proton if the time difference $\Delta t=t_{DC}-t_{TOF}$, corrected
for the event start time, is within a time window $-2\,\,\mathrm{ns}$ to $7\,\,\mathrm{ns}$.

A vertex cut is applied to ensure that the interaction took place within the volume of the
target. 
The electron was required to have a vertex
$-2\,\,\mathrm{cm}<Z_{el}<1.5\,\,\mathrm{cm}$ while the proton vertex cut was set
to $-2.5\,\,\mathrm{cm}<Z_{pr}<2\,\,\mathrm{cm}$ (the target extends from -2.5 cm to
2.5 cm). Additionally the vertex difference between $Z_{el}$ and
$Z_{pr}$ was required to be less than $1.4\,\,\mathrm{cm}$ to reduce
the background from accidental coincidences.

\subsection{Kinematic Corrections}

The geometrical and structural complexity of CLAS is responsible for
minor discrepancies in the measurement of the momentum and direction of
a particle. These discrepancies are thought to be primarily due to
the uncertainty in the magnetic field map and DC position. The effect
of a displacement of the drift chambers and possible discrepancies
in the measured magnetic field on the measured scattering angle $\theta_{rec}$
and momentum $p$ can be parameterized.

The correction function contains 8 parameters describing the drift
chamber displacements and rotations and 8 parameters describing the
possible uncertainties in the magnitude of the magnetic field on the
path of the particle. These parameters can be determined 
using multi-particle exclusive
reactions which are fully contained within the CLAS acceptance. 
In an exclusive reaction all of the products of the reactions
are detected and no mass is missing. Therefore, the kinematics of
the reaction are fully defined and the goodness of fit can be evaluated
using momentum and energy conservation. More details on this method
can be found in Ref. \cite{MomCor}.

For low-energy protons ($P<0.75\,\,\mathrm{GeV/c}$) 
energy loss in the target and detector is significant and needs to be corrected for. 
This energy loss was studied
with the CLAS GEANT simulation and an appropriate correction was applied
to the data.

\subsection{Backgrounds}

Even after the ID cuts described above, pions remain a non-negligible
background in the electron spectrum. Their contribution needs
to be estimated and appropriate corrections applied to the data. This
was done using a sample of pions within EC cuts of $E_{inner}<0.05$ GeV
and $E_{total}<0.1$ GeV. The spectrum of photoelectrons in the Cherenkov
Counters of this pion sample was scaled such that the sum of the
normalized spectrum and that of a ``perfect'' electron sample (from
a simulation normalized to data within a tight EC cut) agreed with
the measured Cherenkov spectrum for electron candidates 
within our regular EC cuts.
This normalized
pion spectrum was then integrated above the software ID cuts of 2.5
and 1.0 photoelectrons (depending on the data momentum range) and
used to estimate the fraction of pions remaining in the electron sample
after the Cherenkov ID cut. This fraction was fit to an exponential in pion
energy and the resulting estimate of the pion contamination (ranging to no
more than 6\%) was used to correct the extracted data. 

A similar technique was used to measure the rate of positrons relative
to that of electrons, by taking positive charge tracks and fitting
their energy spectrum in the EC with a combination of ``pure'' pions
(based on Cherenkov response) and ``golden electrons'' (very high
Cherenkov cut). This positron to electron ratio can be used to estimate
the fraction of the detected electrons which were not scattered from the beam
but came from pair production
$\gamma\rightarrow e^{+}e^{-}$ or the Dalitz decay 
$\pi^{\mathrm{0}}\rightarrow\gamma e^{+}e^{-}$. Once again, an exponential
fit to the ratio was used to estimate this contamination for all kinematic bins
and correct our final data accordingly.

\begin{figure}
\includegraphics[%
  scale=0.45]{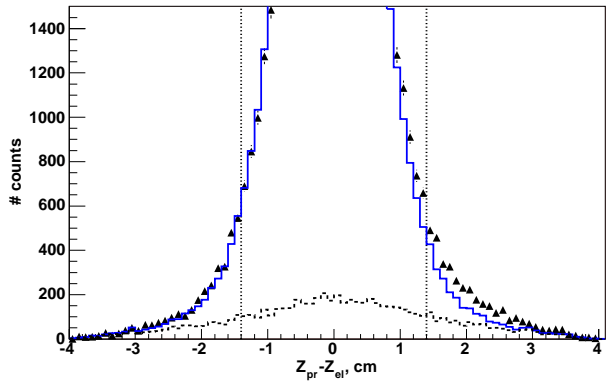}
\caption{\label{fig:acc} (Color online) Data for the difference between the electron
and proton vertex (triangles) compared to a fit (solid histogram) composed
of a simulation of true coincidences (not shown) and measured
accidental coincidences (dash dotted histogram). The vertical dashed lines
indicate the cut used to select data for analysis.}
\end{figure}

Despite the vertex cuts there is still a chance of having an accidental
coincidence between an electron and a proton in the data sample. The
background of accidentals has to be estimated and subtracted.
At the same time, the loss of {}``true'' protons due to
the time and vertex cuts has to be determined. A purely accidental proton
was defined as a positively charged particle with the time-of-flight
measured by the TOF to be at least 12 ns longer than the expected
time-of-flight of a proton with that momentum. The time window for
the accidental proton was taken to be 9 ns, the same as the proton
ID time window, so that the expected arrival time for the accidental
proton would not be more than 21 ns different from the expected arrival
time of the real proton. In the case where the time window of accidentals
is less than 5 ns away from when the deuteron (from elastic scattering
events) would have arrived at the TOF counter, the accidental proton
is defined to be within a 9 ns window starting at  5 ns
after the expected arrival time of a deuterium ion. The average
background of accidental coincidences per nanosecond of the proton
time vertex was calculated from the rate in the {}``accidental
time window'' described above and compared with the unbiased data sample
of coincidences with good proton PID. The level of understanding of
accidentals was also tested using the simulation results. The sum
of the measured accidentals and the simulation is in agreement with the data
on good electron-proton coincidences as selected by PID cuts (Fig.
\ref{fig:acc}). A small discrepancy on the positive side of the $\Delta Z$
distribution is due to another type of unwanted coincidences where
a particle originating from the first electron vertex reinteracts further
along the target cell, liberating a (backward) proton which
arrives on-time with respect to the TOF. Protons produced in such a way enhance
the positive side of the vertex difference distribution. The selected
sample of accidentals contains only off-time events, and therefore does
not fully reproduce the shape of the vertex difference distribution.
A properly scaled sample of these excess events was added to the
sample of purely accidental coincidences defined
using off-time protons.

\subsection{Simulation}

To extract absolute results from our experimental data, the detector
acceptance has to be evaluated and an appropriate correction applied
to the data. An idealized model of all the detector systems of CLAS is implemented
in the code known as {}``GSIM''. The program is built
on the foundation of the GEANT simulation software package, supported
by CERN. GSIM allows simulation of the detector response to a propagating
particle, simulating energy loss as well as emission of secondary
particles during the passage of the particle through parts
of the detector. After the response of the ideal detector is simulated,
existing detector inefficiencies are introduced. This is done using
a separate program called {}``GPP'' (GSIM post-processor). GPP uses
precompiled information on dead regions of the DC and TOF to remove the
signal for these parts of CLAS from the GSIM output. The final output
is then analyzed exactly the same way as the real data.

The events used as input for the CLAS GSIM simulation were generated
following the cross section Eq.~\ref{eq:ccoffshell}. The Paris wave function~\cite{Paris}
was used to select the momentum of the ``spectator'' nucleon first. A comparison with the
Argonne V18 wave function~\cite{Argonne} showed a negligible difference in the
momentum distributions. The generated nucleon momentum can either be directly used
following the prescription for the non-relativistic spectral function (Eqs.~\ref{eq:pps},\ref{eq:salphas}) or
as the ``internal momentum'' in the light cone description, Eqs.~\ref{eq:sapt}--\ref{eq:SpFun}.
From the spectator nucleon kinematics,
we then calculate the initial four-momentum of the struck nucleon and determine the
scattered electron kinematics in the rest frame of that nucleon, then transform it
back to the lab frame. That way, all of the ``starred'' variables in Eq.~\ref{eq:ccoffshell}
are automatically evaluated with the proper relativistic rescaling. 

The electron scattering
cross section used to generate the electron kinematics is based  on the code RCSLACPOL
that was developed at SLAC \cite{Abe:1998wq}. It uses
parametrizations of world data on unpolarized structure functions and
elastic form factors. These parametrizations
are described in~\cite{E155Q2} and
are based on fits to unpolarized structure function data from
NMC~\cite{NMCF2} and SLAC~\cite{BodekSF, E140, NE11delta, E143R}.
 The nucleon form factors
were taken from Ref.~\cite{Bosted}. All form factors and structure functions
for bound nucleons are assumed to be equal to the free ones at the corresponding
values of $x$ (in the DIS region) or $W$ (in the resonance region, with a smooth
transition between both). The free neutron structure function $F_{2n}$ was extracted
from fits to the world data on the deuteron in a self-consistent manner by ensuring that our model,
integrated over all spectator kinematics and summed over both proton and neutron 
contributions to electron scattering, agrees with those fits. 

Three different versions
of the code were compiled to satisfy our needs for simulation of electron
scattering on $^{2}\mathrm{H}$: 
1) elastic scattering on one nucleon in the deuteron (with the other being a spectator), 
including the elastic radiative tail; 
2) inelastic scattering on one nucleon
in the deuteron (with and without radiative corrections); and
3) elastic scattering off the deuteron nucleus as a whole.
Radiative effects can be included in the simulation following the 
prescription by Mo and Tsai~\cite{MoTsai}. In the first two cases, these radiative corrections are applied
to the electron scattering cross section for the struck nucleon in its rest frame, while the
spectator simply determines the kinematic transformation into the lab system.
The generator is capable of simulating both inclusive \dee \  
(by adding the first two processes for both protons and neutrons with the third one) and semi-inclusive
\deeps \    processes, which is controlled by a configuration file. 
While this generator may not be very realistic in its description of the underlying physical
processes (since it
does not contain FSI, non-nucleonic currents in deuterium, or modifications
of the nucleon structure function for off--shell nucleons), it is sufficiently accurate (see below) to
allow a largely unbiased extraction of the acceptance and efficiency of CLAS, by comparing
accepted simulated events to the initial distribution of generated events.

The quality of the simulation procedures can be evaluated by comparing
the predicted number of counts for well-studied processes in data and
simulation. To date, one of the best studied cross sections in nuclear
physics is that of elastic electron scattering from a free proton.
To select elastic events a cut on the invariant mass $W$ was used:
$0.9<W<1.1\,\,\mathrm{GeV}$. The overall shape is reproduced well
and the measured cross section lies well within 10\% of the simulated
one at low $Q^{2}$ (where our statistical error allows a significant comparison). 
The $Q^{2}$ distribution of the simulated inclusive
cross section for quasi-elastic scattering on deuterium is also in
good agreement with the experimental data. Here the events were also
selected using the invariant mass cut $0.9<W<1.1$ GeV. In the region
of relatively good statistics at low $Q^{2}$ the deviation from unity
on the data to simulation ratio does not exceed 10\%. Finally,
the rate of inclusive \dee $X$ events for all final state invariant
masses $W$ agrees with the prediction of our model to within 5--10\%.

A sample of simulated events that exceeds the statistics of the experimental
data by a factor of 10 was generated for the \deeps \   reaction
and was used in the analysis to correct the data for detector acceptance
and bin averaging effects. The high event count of the Monte Carlo
assures that the statistical error of the data points are not dominated
by the statistical error of the simulation.

\subsection{Result Extraction}

The events from the data set were sorted in four-dimensional
kinematic bins in $W^*$ (or $x^*$), $Q^2$, $p_s$ and 
$\cos\theta_{pq}$ (or $\alpha_s$ and $p_T$). We chose two
bins in $Q^2$, one with $1.2$ (GeV/c)$^2 \leq Q^2 \leq 2.1$ (GeV/c)$^2$
(average $Q^2 = 1.8$ (GeV/c)$^2$) and one with 
$2.1$ (GeV/c)$^2 \leq Q^2 \leq 5.0$ (GeV/c)$^2$
(average $Q^2 = 2.8$ (GeV/c)$^2$, and five bins in $p_s$, with average
values of $p_s = 0.3, 0.34, 0.39, 0.46$ and $0.53$ GeV/c.

To extract the final results, the above bins were filled
separately for the following categories of events: 1) experimental
data with all the standard electron and proton ID cuts; 2) accidental
electron-proton coincidences based on experimental data; 
3) coincidences with protons from secondary scattering events;
4) simulated data for the elastic scattering
on a bound neutron, including the radiative elastic tail; 
5) simulated data for the inelastic scattering
on a bound neutron. Accidental coincidences and coincidences with
secondary protons were then subtracted from the data on a bin-by-bin
basis. The simulated elastic scattering data were also used to subtract
the elastic radiative tail from the experimental data. For this purpose
both data and simulation were first integrated in the range of the
invariant mass of the unobserved final state $W^{*}$ from 0.5 to
1.1 GeV. The elastic radiative tail in the simulation was then scaled
by the ratio of the data to the simulation and subtracted. 

As was previously discussed, in the spectator picture, the cross section
for the off-shell nucleon can be factorized as a product of the bound
nucleon structure function and the nuclear spectral function, multiplied
by a kinematic factor (see Eq.~\ref{eq:ccoffshell}). Using the
data of this experiment, it is possible to extract this product, and,
in the region where FSI are small and the spectral function is well
described by the model, even the off-shell structure function by itself.
To do that, the experimental data (with accidentals, rescattered proton
events, and elastic radiative tail subtracted) were first divided
by the simulated inelastic data. The simulated events were generated using the
cross section Eq.~\ref{eq:ccoffshell} with full consideration of radiative
effects. To extract the product of structure and spectral functions,
the ratio of data to simulation was multiplied with the product 
$F_{2n}(x^{*},Q^{2})\times S(\alpha_s, p_{T})$,
calculated using the same model that was used in the generator. 
Similarly, to obtain the product of the structure function $F_{2n}$
with the probability distribution for the proton momentum in deuterium,
we multiplied the ratio  of data to simulation with the factor
$F_{2n}(x^{*},Q^{2})\times P(\vec{p}_{s})$ from our generator model.
In both cases,
the dependence of the extracted data on the specific model for the
simulation is minimized, since the {}``input'' ($F_{2n}$ and $S(\alpha_s, p_{T})$
or $P(\vec{p}_{s})$)
cancels to first order. Basically, this procedure
corrects the data for the detector acceptance, bin migration and radiative
effects, and produces a ``normalized cross section'' by 
dividing out the kinematic factor $\frac{4\pi\alpha_{EM}^{2}}{x^{*}Q^{2}}$
as well as the factor in square brackets in Eq.~\ref{eq:ccoffshell}
(which depends weakly on the ratio $R=\sigma_{L}/\sigma_{T}$).
To extract the (``off-shell'') structure function  \f2neff, the
ratio of data to simulation was multiplied with the free nucleon structure
function $F_{2n}(x^{*},Q^{2})$. This assumes that the spectral function
used in the simulation describes the momentum distribution of the
spectator protons reasonably well.

\subsection{Systematic Uncertainties}

To simplify the statistical error calculation, all the corrections
for the detector inefficiencies and data sample contamination (except
for accidentals and the radiative elastic tail) were applied to the
simulated events.

The efficiency of the CC electron ID cut is well reproduced in the
simulation. A 1\% systematic uncertainty enters here to account for
the observed deviation of the cut efficiency from sector to sector.
The EC ID cut efficiency is reproduced only partially. The efficiency
of the cut in data was found to be 95\%, however the same cut, applied
to the simulation, is 98\% efficient. The difference might be a result
of data being contaminated with pions, despite the increased CC threshold.
The simulated data were scaled down by a constant factor of 0.97 to
account for the difference in the effect of the cut. A 2\% systematic
uncertainty was assigned to this factor due to the uncertainty about
the source of the deviation. A variable factor that ranges from 1.06
to less than 1.01 was used to introduce pion contamination into the
simulation. The factor varies with the particle scattering angle and
momentum. A variable factor was also applied to the electron spectrum
in the simulation to introduce electrons coming from electron-positron
pair creation. The resulting systematic uncertainty was estimated
by varying these factors by 50\% of their deviation from unity. The
resulting change in the distribution in each of the final histograms
was used as an estimate of the systematic uncertainty of these corrections.

Some additional corrections were applied to the proton spectrum. A
constant factor of 0.99 was introduced to reflect the difference in
the effect of the proton timing ID cut on the real versus the simulated data. 
The systematic uncertainty
of 0.5\% on this number accounts for the momentum dependence of the effect.
A factor dependent on the proton momentum was applied to the simulated
data to account for the discrepancy between data and simulation in
the effect of the cut that was set on the difference between the electron
and proton vertices. The systematic uncertainty here is evaluated individually
for each histogram, by varying the correction by 50\%.

A major contribution to our systematic error comes from remaining differences
between the simulated and the ``true'' inefficiencies of CLAS. Even after
removing bad channels and accounting for all known detector problems, we find
that the ratio of simulated to measured rates for reconstructed protons varies
from sector to sector. We use the RMS variation between sectors to estimate
this systematic error as about 11\% on average. We also include a 3\% scale
error on the target density, effective target length, and beam charge calibration.

The data were corrected for the radiative elastic tail and accidental 
coincidences by direct subtraction of normalized (simulated or real) data
(see previous subsection). The
normalization factors were varied by 50\% of their deviation from unity
to estimate the systematic errors due to these corrections. The uncertainty
on the inelastic radiative corrections was also calculated as 50\% of
the deviation from unity of the correction factor. We checked our radiative
correction procedure against the existing code ``EXCLURAD''~\cite{Exclurad}
for the case of
quasi-elastic scattering (pn final state) and found good agreement within the
stated uncertainties.

A final systematic uncertainty comes from the model dependence of our simulated data.
While the model input cancels in our extracted values for 
$F_{2n}(x^{*},Q^{2})\times S(\alpha_s, p_{T})$ to first order, both migration between
adjacent kinematic bins and distribution of events within a bin (where the CLAS 
acceptance might vary) are somewhat model-dependent. We estimated this effect
by modifying the model input to agree with the cross section extracted from our data.
The deviation of the simulated events with this modified cross section from the
data is a direct measure of the magnitude of this systematic error. We found its magnitude
to be generally below 5\%, going up to 10\% for higher proton momenta.

All systematic errors were added in quadrature and are shown as shaded bands
in the Figures in the following section.
The summary of systematic uncertainties is presented in Table \ref{tab:errors}. 

\begin{table}
\begin{center}\begin{tabular}{|p{0.5\columnwidth}|p{0.5\columnwidth}|}
\hline 
Source of Uncertainty&
Typical Range (in \% of data value)\tabularnewline
\hline
\hline 
EC ID Cut&
2\tabularnewline
\hline 
Trigger Efficiency&
2\tabularnewline
\hline 
Secondary Electrons&
0.7\tabularnewline
\hline 
Electron Vertex ID Cut&
0.6\tabularnewline
\hline 
Proton Timing ID Cut&
0.5\tabularnewline
\hline
CC Efficiency&
1\tabularnewline
\hline
Pion Contamination&
0.5 ... 3\tabularnewline
\hline
$e^{+}/e^{-}$ Contamination&
0 ... 0.75\tabularnewline
\hline
Pure Accidental Coincidences&
0 ... <1.2> ... 4\tabularnewline
\hline
Coincidences with Knock-out Proton &
0 ... <2.3> ... 6 \tabularnewline
\hline
Vertex Difference Cut&
0.75 ... 1.5\tabularnewline
\hline
Quasi-elastic Radiative Corrections&
0 ... <1.9> ... 11\tabularnewline
\hline
Inelastic Radiative Effects&
0 ... <2.7> ... 12\tabularnewline
\hline
Luminosity&
3\tabularnewline
\hline
Tracking Inefficiency&
11\tabularnewline
\hline
Bin Migration \& Model-Dependence of Acceptance&
0 ... <5.2> ... 10\tabularnewline
\hline
\hline 
Total&
15.5 ... <16.9> ... 34.1\tabularnewline
\hline
\end{tabular}\end{center}

\caption{\label{tab:errors} Systematic errors in percent of the data values.
The typical range of the error as well as their RMS values (in brackets) are given.}
\end{table}



\section{Results}

\begin{figure}
\begin{flushleft}\includegraphics[%
  scale=0.42]{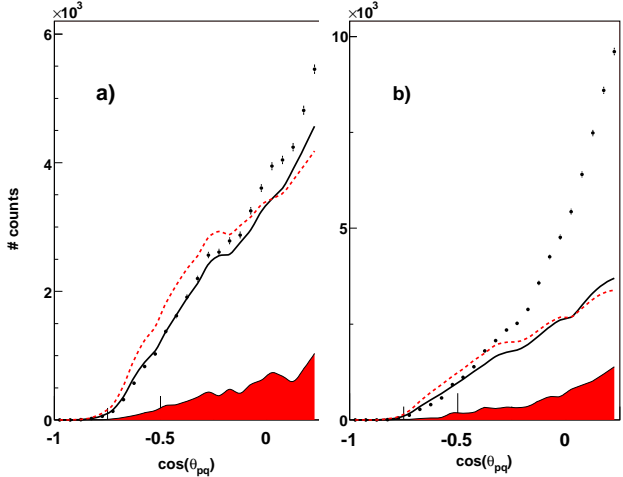}\end{flushleft}
\caption{\label{fig:cthpq_ps} (Color online) Data (points) and results of the Monte Carlo (MC)
simulation based on two different PWIA models (solid and
dashed curves) for the total number of counts  versus 
$\cos\theta_{pq}$ for proton momenta 
$p_{s}=280 $--$ 320\,\,\mathrm{MeV/c}$
(a) and $p_{s}=360 $--$ 420\,\,\mathrm{MeV/c}$ (b), 
integrated over electron
kinematics. The total systematic error is indicated by the shaded band.}
\end{figure}

In the following, we show several representative histograms (one--dimensional
projections of the four--dimensional bins), comparing our data to our simple
PWIA spectator model to elucidate some general trends. 

\begin{figure}
\begin{flushleft}\includegraphics[scale=0.42]
{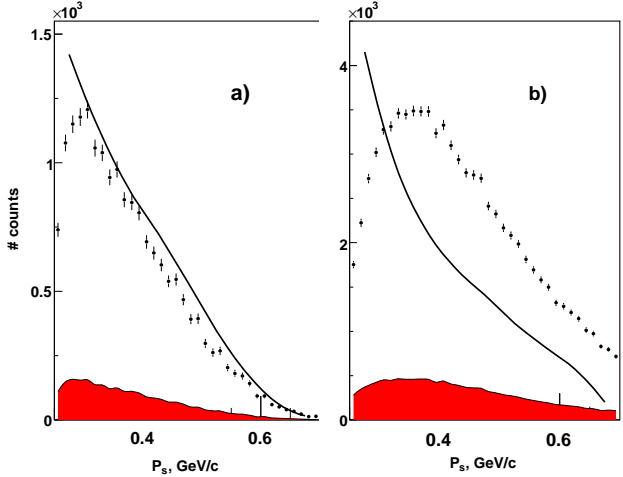}\end{flushleft}
\caption{\label{fig:ps_w} (Color online) Momentum distribution of the recoiling proton. Data
(points) are compared with our MC simulation (solid curve) for the range of
recoil angle $-1.0<\cos\theta_{pq} <-0.3$ (a) and $-0.3<\cos\theta_{pq} <0.3$
(b). All events within a missing mass range $1.1<W^{*}<2.0\,\,\mathrm{GeV}$
were summed together for this plot.}
\end{figure}

\begin{figure*}
\includegraphics[scale=0.67]{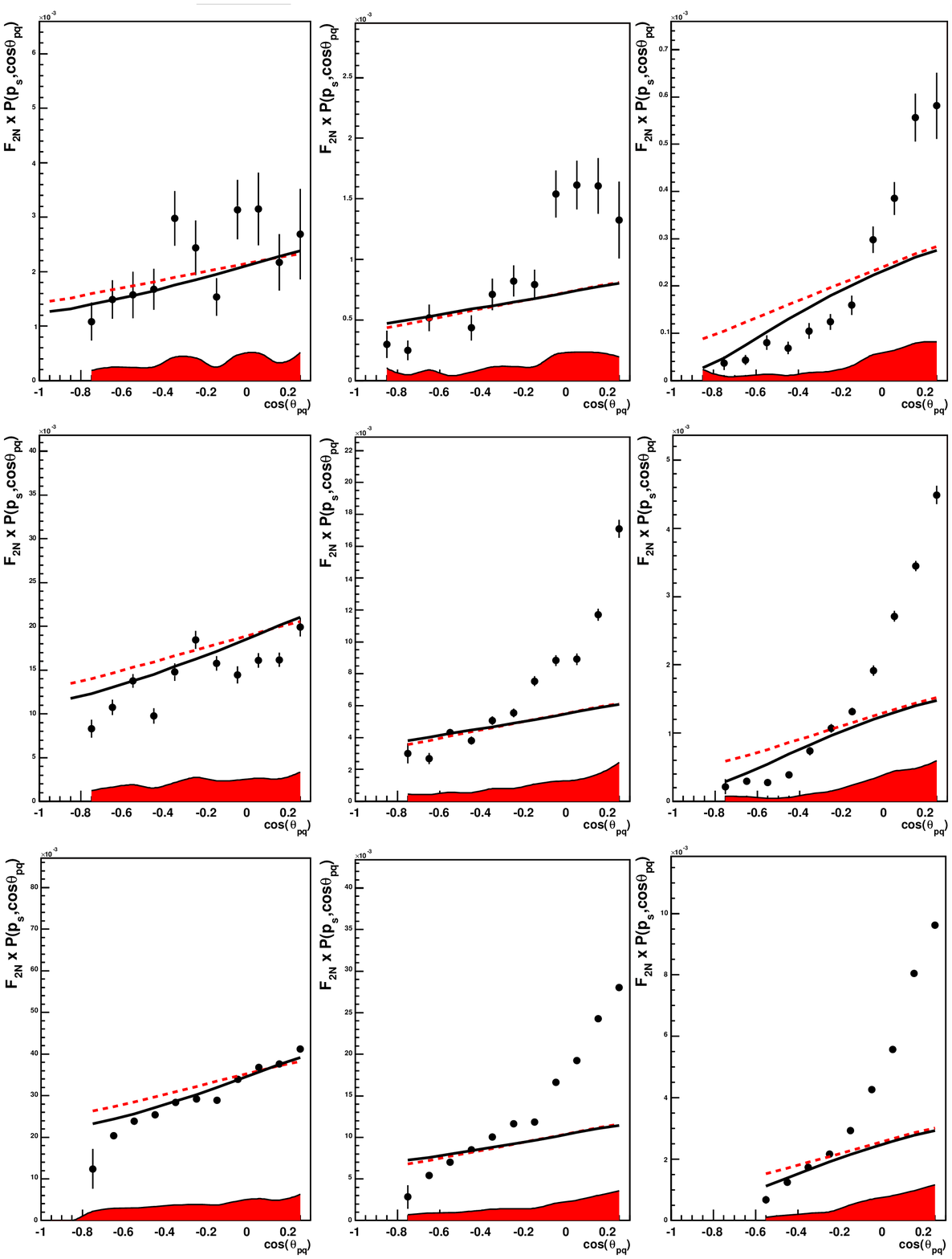}
\caption{\label{fig:cthpq_w_pub} (Color online) 
Results for the normalized cross section (equivalent to the product
$F_{2n}\times P(\vec{p}_{s})$ in the spectator picture) for the reaction
\deeps $X$. Each row is for a different mass $W^*$ of the unobserved
final state $X$, namely $W^* = 0.94$ GeV (quasi-elastic scattering) in
the first row, $W^* = 1.5$ GeV in the second and $W^* = 2$ GeV in
the third. The three columns are for three different proton momentum
ranges, with average momenta of $p_s = 0.3$, $0.39$ and $0.56$ GeV/c, 
respectively.
All data (filled circles with statistical error bars) 
are for our lower $Q^2$ bin (with average $Q^2$
of 1.8 (GeV/c)$^2$). The two lines come from our simple PWIA spectator
model using a light-cone wave function (solid line) or a non-relativistic
WF (dashed line), while the shaded band at the bottom indicates the
systematic error.}
\end{figure*}

In Fig. \ref{fig:cthpq_ps} we show as a first step the
accumulated number of protons (in coincidence with a scattered electron)
 for several bins in $\cos\theta_{pq}$, where $\theta_{pq}$ is the
angle  between the virtual
exchanged photon and the proton. The data are not corrected for
acceptance and efficiency and therefore fall off at large angles
where CLAS has limited acceptance.
The curves shown are from our simulation of these data,
including the CLAS acceptance and
without any normalization.
Using the light cone
prescription (Eq.~\ref{eq:SpFun}) for the momentum distribution 
of the initial proton (solid curve), good agreement between the data and 
our Monte Carlo (MC) simulation is observed up to $\cos\theta_{pq} \approx-0.3$. The result
for the non-relativistic wave function (Eq.~\ref{eq:pps}, dashed line) is 
similar in these kinematics.
At more forward angles the data exceed the simulation by a large factor,
especially at higher momenta (Fig.~\ref{fig:cthpq_ps}b), indicating a breakdown
of the pure PWIA spectator picture. We assume that this enhancement is due to FSI
between the struck neutron and the spectator proton (see below).

The momentum distribution plotted separately for backward ($\theta_{pq}>108^{\circ}$) 
and transverse 
($72^{\circ}<\theta_{pq}<108^{\circ}$)
proton kinematics confirms
this picture for the relative importance of non-PWIA processes (Fig. \ref{fig:ps_w}). 
The momentum distribution of the backward
protons is reasonably well described by the PWIA model, indicating that distortions
due to FSI
are rather small in this region. At the same time, the momentum distribution
for the transverse protons is strongly enhanced at momenta above 300
MeV/c, as predicted by several models of 
FSI~\cite{Melnitchouk:1996vp,Frankfurt:1996xx,CiofidegliAtti:2003pb,Lagetquel}.
For momenta below about 300 MeV/c, the acceptance and efficiency of CLAS
for protons falls off even faster than predicted by our Monte Carlo simulation. This explains
the fall-off at low momenta in Fig.~\ref{fig:ps_w}.

In Fig. \ref{fig:cthpq_w_pub} we look at the angular distribution of the protons
in more detail. The reduced cross section described in the previous section
is plotted for three different proton momenta (increasing from left to right), 
as well as three different
missing mass ranges of the unobserved final state (increasing from top to bottom)
in the reaction 
\deeps $X$. Several trends can be observed:
\begin{itemize}
\item{
At proton momenta
around 300 MeV/c, the extracted reduced cross section is consistent with 
our simple PWIA spectator model throughout the whole
angular range and for all final state  masses. 
This is consistent with expectations that destructive and
constructive interference effects between FSI and PWIA cancel 
roughly in this
momentum range~\cite{Frankfurt:1996xx, Lagetquel}.}
\item{
For larger proton momenta, deviations from PWIA behavior show up
as an increase of the normalized cross section at transverse kinematics. 
This increase appears approximately
around $\cos\theta_{pq} = -0.3$ and continues beyond 
$\cos\theta_{pq} = 0$ ($\theta_{pq} = 90^{\circ}$). 
Such an increase is not likely due to uncertainties in the deuteron wave function,
which is isotropic in the non-relativistic case and is equal to the non-relativistic
wave function for transverse proton momenta if one uses light-cone wave functions.
However, such an effect is expected
within models of FSI
due to the initial motion of the nucleon on which the rescattering
occurs (see Fig. 3 in Ref.~\cite{Frankfurt:1996xx}  and Ref.~\cite{Lagetquel}).
The strength of FSI in these models is the largest for the highest
recoiling proton momenta, consistent with the trend of the data. }
\item{The non-PWIA effects seem to be more pronounced for the largest missing masses
(see also below).
This behavior is in qualitative
agreement with the FSI model by Cioffi delgi Atti and 
collaborators~\cite{CiofidegliAtti:2003pb,CiofiEPJA17},
where the strength of rescattering is related to the 
number of hadrons in the final state (Eq.\ref{eq:FSIcioffi}).}
\end{itemize}

\begin{figure}
\includegraphics[%
  scale=0.45]{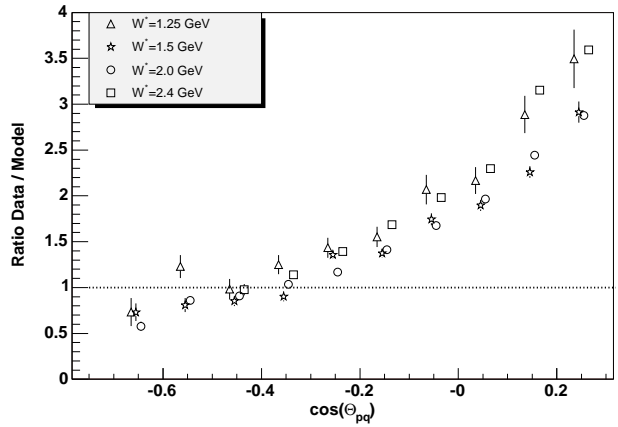}
\caption{\label{fig:cthpq_rat_w2}Ratio of data to model as a function of
$\cos\theta_{pq}$ for four values of missing mass $W^{*}$ at
$p_{s}=460\,\,\mathrm{MeV/c}$ and $Q^{2}=1.8\,\,\mathrm{GeV^{2}/c^{2}}$.}
\end{figure}

This last point can be seen more clearly in Fig. \ref{fig:cthpq_rat_w2}
which shows the ratio between the observed cross section and the 
prediction of our PWIA spectator model for proton momenta around
0.46 GeV/c, for four different ranges in final-state missing mass (slightly
offset from each other for each point in $\cos\theta_{pq}$). The data for
different missing mass values are statistically close to each other
(and close to unity) in the backward
region where rescattering effects can be assumed to be small.
Conversely, in transverse kinematics
the ratio substantially exceeds one and is largest for the highest $W^{*}$ bin. 
The enhancement in transverse kinematics is also large in the 
$\Delta-$resonance region.
This could be due to $\Delta$--production in
FSI between the struck neutron and the ``spectator'' proton.

%

\begin{figure}
\begin{flushleft}\includegraphics[%
  scale=0.42]{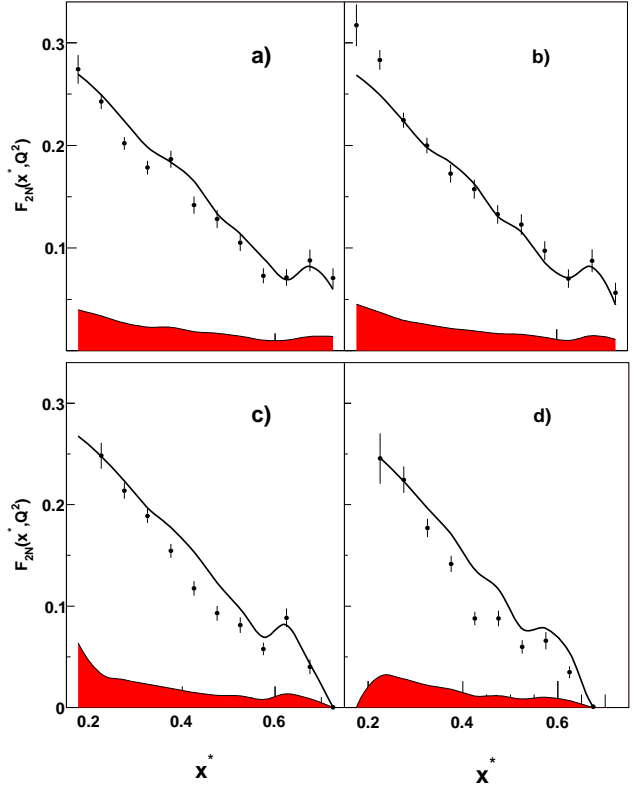}\end{flushleft}
\caption{\label{fig:F2n} (Color online) Results for the extracted {}``off-shell'' structure
function $F_{2n}^{\mbox{\tiny eff}}$ of the neutron in the PWIA spectator picture. The
model (solid curve) is a simple parameterization of the free on-shell
neutron structure function, modified to account for the kinematic
shift due to the motion of the {}``off-shell'' neutron. The sections
of the plot correspond to different recoiling proton momenta: $p_{s}=300\,\,\mathrm{MeV/c}$
(a), $p_{s}=340\,\,\mathrm{MeV/c}$ (b), $p_{s}=460\,\,\mathrm{MeV/c}$
(c) and $p_{s}=560\,\,\mathrm{MeV/c}$ (d). 
The quantity plotted here is similar (but not identical) to the quantity $F^{(s.i.)}$
defined in the paper by Simula~\protect{\cite{Simula:1996xk}}. }
\end{figure}

Concluding that the spectator PWIA model works reasonably well in the region
of large backward angles ($\cos\theta_{pq} <-0.3$), we concentrate on this
region to study the momentum (off-shell) dependence of the effective 
electron scattering cross section on the bound neutron. 
%
At first, we directly compare the extracted effective
structure function  of the off--shell neutron,  \f2neff, 
for inelastic final states ($W^* > 1.1$ GeV)
to the on--shell structure function (see  Fig.~\ref{fig:F2n}).
To obtain this structure function, 
the measured cross section was divided by the proton momentum distribution,
Mott cross section and the kinematic factor as explained in the previous section. 
Even within the PWIA picture, the results could have a $p_s$--
dependent scale error because our simple model may not
describe the nucleon momentum distribution in deuterium perfectly;
however, the $x^*$--dependence in each individual panel would be largely unaffected by
such a scale error.
Indeed, the
data agree reasonably well with the simple parameterization of the free neutron structure
function from our model at the two lower momenta (with average deviations of $\pm 10\%$).
At the higher two momenta,
the data fall below the model in the range of $x$ between 0.3 and
0.6 by as much as $20\%$ -- $30\%$. 
Such a reduction in the structure function is expected in several models
of modification of bound nucleon structure~\cite{Melnitchouk:1996vp}. 
Some residual FSI
might also contribute to the observed $x^*$--dependence, for instance by enhancing
the region of small $x^*$ (corresponding to large $W^*$).

\begin{figure}
\begin{flushleft}\includegraphics[scale=0.5]{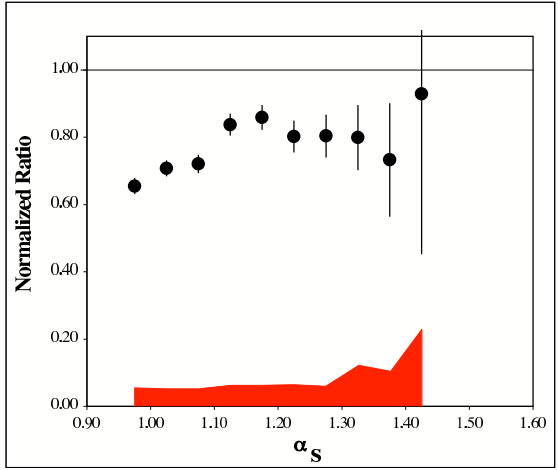}\end{flushleft}
\caption{\label{fig:F2nrat} (Color online) 
Ratio of the extracted {}``off-shell'' structure
function $F_{2n}$ at $x = 0.55, Q^2 = 2.8$ (GeV/c)$^2$ to that at 
$x = 0.25, Q^2 = 1.8$ (GeV/c)$^2$, divided by the ratio of the free structure functions
at these kinematic points. The error bars are statistical only and the shaded band indicates
the overall systematic error. This plot is for similar (but not identical) kinematics as
Fig. 6 in the paper by Melnitchouk et al.~\protect{\cite{Melnitchouk:1996vp}}.}
\end{figure}

To reduce the model dependence of such comparisons as in Fig.~\ref{fig:F2n},
the authors of Ref.~\cite{Melnitchouk:1996vp} suggested to take the ratio between
the extracted ``off-shell'' structure function at some relatively large value of $x^*$
(where most models predict the biggest off-shell effects) to that at a smaller value
of $x^*$ where the EMC--effect is known to be small. 
This ratio (normalized to the same ratio for the free neutron
structure function, $F_{2n}$) is plotted in Fig.~\ref{fig:F2nrat}
for a range of transverse momenta 0.25 GeV/c $\le p_T \le$ 0.35 GeV/c. Nearly all dependence
on our model cancels in this ratio;
only the overall scale depends on the ratio of $F_{2n}$ for {\em free} neutrons
at two different values of $x$, which is not perfectly well known.
The ratio plotted in Fig.~\ref{fig:F2n} is also independent of the
deuteron momentum distribution $P(\vec{p}_{s})$; however,
according to some models~\cite{CiofidegliAtti:2003pb}, FSI effects could be 
different for different $x^*$. This seems to be born out by  Fig.~\ref{fig:F2nrat}: While
all PWIA models of off-shell effects predict unity for the ratio at values of the  light cone
variable $\alpha_s$ around 1, we find a strong suppression in the region up to 
$\alpha_s \approx 1.1$ (corresponding to $\theta_{pq}$ around $90^{\circ}$)
where FSI are most pronounced. This behavior could be explained within the
FSI model of Ref.~\cite{CiofidegliAtti:2003pb} which predicts larger
FSI effects for final states with a larger number of hadrons, leading to an increase of
the denominator (cross section at small $x^*$, which corresponds to large energy
transfer to the unobserved final state).

Beyond $\alpha_s \approx 1.1$, the data still lie below unity (by about
17\%) but appear fairly constant with $\alpha_s$. Although this suppression could
be interpreted as an off--shell effect, the data appear inconsistent with some of
the more dramatic predictions of a steep falloff for the ratio at high $\alpha_s$
({\it e.g.}, Ref.~\cite{Frankfurt:1981mk}). The prediction for this ratio from the 
6-quark cluster model~\cite{Carlson:1994ga} varies between 0.7 and 1 at $\alpha_s = 1.4$ and
is therefore compatible with our result.
Once realistic calculations including FSI effects become available for the
kinematics of our data set, a more quantitative comparison with various
models for the off--shell behavior of the structure function $F_2(x^*,Q^2, p_s)$
will be feasible. Such calculations are underway~\cite{CiofiPC, Lagetquel}.

\section{Summary}

Taking advantage of the large solid angle acceptance of the CEBAF
Large Acceptance Spectrometer, a large amount 
of data ($\approx$ 350K events) was collected
on the reaction \deeps $X$ in the exotic region of extreme backward
proton kinematics. The data range from 1.2 to 5 (GeV/c)$^2$
in momentum transfer $Q^{2}$ and reach values of the missing mass
of the unobserved final state $W^{*}$ of up to $2.7\,\,\mathrm{GeV}$.
Protons with momentum $p_{s}$ as low as $280\,\,\mathrm{MeV/c}$
and up to $700\,\,\mathrm{MeV/c}$ were detected, at angles $\theta_{pq}$ relative
to the direction of the momentum transfer extending up to more than $140^{\circ}$.
In terms of the light cone variables, the data span values of the
light-cone fraction $\alpha_{s}$ up to about 1.7, with a minimum
proton transverse momentum relative to $\hat{q}$ of $150\,\,\mathrm{MeV/c}$
and up to $600\,\,\mathrm{MeV/c}$.

Reduced cross sections were extracted as a function of
$W^{*}$ (or Bjorken--variable $x^{*}$) and $\alpha_{T}$, $\vec{p}_{T}$
(or $\cos\theta_{pq}$, $p_s$), for two large bins in $Q^2$,
allowing us to test theoretical calculations against the presented
data. Comparison with a simple PWIA spectator model shows 
moderately good agreement in
 the kinematic
region of lower momenta and $\cos\theta_{pq} < -0.3$. 
For increasing ``spectator'' momenta $p_{s} > 0.3$EG V/c
FSI and other non-PWIA effects become strong, especially
in the region of proton scattering angles $\cos\theta_{pq} > -0.3$.
These effects seem to depend on the invariant mass $W^{*}$; on the other hand,
no strong dependence of these effects on momentum transfer
$Q^{2}$ is observed. This behavior is in qualitative agreement with
models~\cite{CiofidegliAtti:2003pb,CiofiEPJA17} that describe the strength of FSI in terms of the number
of hadrons in the final state $X$. The angular $(\theta_{pq})$ and momentum
$(p_{s})$ dependence of the observed strength in the cross section
in the quasi-elastic region (where $X$ is a neutron in its ground state)
are also in good agreement with detailed calculations~\cite{Lagetquel}  showing
a transition from destructive interference below $p_{s}=300\,\,\mathrm{MeV/c}$
to a strong enhancement at $p_{s}>400\,\,\mathrm{MeV/c}$ around $\cos\theta_{pq} =0.2$
(see Fig.~\ref{fig:cthpq_w_pub} and also Ref.~\cite{Cornel}). 

A depletion compared to the PWIA model is observed in
the data at $\cos\theta_{pq} <-0.3$ and for high $p_{s}$, where
the struck neutron is far off its mass shell. This reduction
might be 
due to nucleon structure modifications. It is especially apparent
in the region of moderate $x^{*}$ which overlaps in part with the
nucleon resonance region. However, it is also possible that our
simple model predicts too much strength in the deuteron momentum
distribution at these higher momenta. This would lead to an
``apparent'' depletion for all values
of $x^{*}$ (or $W^{*}$), which would be somewhat modified by a remaining
FSI--induced enhancement at high $W^{*}$. 

Ultimately, our data will serve to constrain 
detailed theoretical calculations, including
off-shell and FSI effects. Once these effects are well-understood at high
spectator momenta, one can safely extract the 
neutron structure function at lower momenta where those 
corrections are smaller and where their uncertainty will not affect the 
result. This method will
be used in the upcoming  ``BoNuS'' experiment at Jefferson Lab.
A statistically improved data set with much larger kinematic coverage
can be obtained once Jefferson Lab has been upgraded to 12 GeV beam
energy.

\begin{acknowledgments}
We would like to acknowledge the outstanding effort of the Accelerator, Target Group, 
and Physics Division staff at TJNAF that made this experiment possible.

This work was supported by the U.S. Department of Energy, 
the Italian Istituto Nazionale di Fisica Nucleare,  the U.S. National 
Science Foundation, the French
 Commissariat \`a l'Energie Atomique,
the French Centre National de la Recherche Scientifique,
 and the Korea Science and Engineering Foundation. 
The Southeastern Universities Research Association (SURA) operates the 
Thomas Jefferson National Accelerator Facility for the United States
Department of Energy under DOE contract DE-AC05-84ER40150 Modification No. M175.
\end{acknowledgments}



\end{document}